\documentclass{article}
\usepackage{spconf,amsmath,amssymb,graphicx,bm,cite}

\newcommand{\refeq}[1]{Eq.(\ref{eq:#1})}
\newcommand{\reftb}[1]{Table \ref{tb:#1}}

\title{ATTENTION-BASED ASR WITH LIGHTWEIGHT AND DYNAMIC CONVOLUTIONS}
\name{Yuya Fujita$^1$, Aswin Shanmugam Subramanian$^2$, Motoi Omachi$^1$, Shinji Watanabe$^2$}
\address{
 $^1$Yahoo Japan Corporation, Tokyo, JAPAN\\
 $^2$Center for Language and Speech Processing, Johns Hopkins University, Baltimore, MD, USA\\
 \{yuyfujit, momachi\}@yahoo-corp.jp, \{aswin, shinjiw\}@jhu.edu
}

\begin{document}
\ninept
\maketitle
\begin{abstract}
End-to-end (E2E) automatic speech recognition (ASR) with sequence-to-sequence models has gained attention because of its simple model training compared with conventional hidden Markov model based ASR. 
Recently, several studies report the state-of-the-art E2E ASR results obtained by Transformer.
Compared to recurrent neural network (RNN) based E2E models, training of Transformer is more efficient and also achieves better performance on various tasks.
However, self-attention used in Transformer requires computation quadratic in its input length.
In this paper, we propose to apply lightweight and dynamic convolution to E2E ASR as an alternative architecture to the self-attention to make the computational order linear.
We also propose joint training with connectionist temporal classification, convolution on the frequency axis, and combination with self-attention. 
With these techniques, the proposed architectures achieve 
better performance than RNN-based E2E model and performance competitive to state-of-the-art Transformer on various ASR benchmarks including noisy/reverberant tasks.
\end{abstract}
\section{Introduction}
\label{sec:intro}
In the research field of automatic speech recognition (ASR), end-to-end (E2E) models are becoming mainstream. There are several advantages compared to conventional hidden Markov model based approaches. 
The conventional approaches include several components such as lexicon, acoustic and language models. 
Each component is trained independently with a different criterion, which does not correspond to globally optimizing the whole ASR system.
Also, some linguistic knowledge is necessary to create a lexicon.
On the other hand, E2E models are composed of a single neural network so there is no need to train multiple components separately or use any linguistic knowledge.

Many efforts have been made \cite{AttentionNIPS2015,LAS2016,amodei2016deep,Prabhavalkar2017,Shinji2017hybrid,Zeyer2018,Luscher2019} to improve the performance of E2E models hence currently state-of-the-art E2E methods are comparable to or sometimes better than conventional methods. 
For example, Google researchers reported their E2E model outperforms conventional methods with their in-house voice search data \cite{Chiu2018google}. Also, with data augmentation, the performance of publicly available corpora 
such as Librispeech \cite{Libri2015} and Switchboard \cite{SB1992}
is better than highly tuned conventional systems \cite{Park2019}.
Those papers described above use a recurrent neural network (RNN) as its key component. 
However, one disadvantage of RNNs is the inefficiency of back propagation computation with GPU compared to non-recurrent models. This is because of the sequential structure of RNNs.

To overcome this disadvantage, Transformer was proposed \cite{TransformerNIPS2017}. 
Transformer does not use any recurrent architecture hence efficient parallelization of the back propagation computation is possible. 
Transformer was originally proposed for neural machine translation and is also applied to ASR in \cite{dong2018speech,Zhou2018}. 
By joint training and decoding with connectionist temporal classification (CTC) \cite{Graves2006CTC}, Transformer's performance is superior to RNN-based models \cite{Karita2019,karita2019comparative}.

Although the parallel computation of Transformer is more efficient than RNNs, the computational complexity of the self-attention layer used in Transformer is quadratic in the input feature length, which requires huge GPU memory.
To make this complexity linear, this paper employs convolution instead of self-attention \cite{gehring2017convolutional}.
However, the number of parameters and computational complexity of regular convolution is quadratic in the dimension of the input feature. 
Depthwise convolution \cite{sifre2014rigid} which performs convolution independently over every dimension is an option to reduce the number of parameters and the computational complexity to linear.
This enables stacking more layers and using a longer kernel length than regular convolution. 

In order to further reduce the number of parameters, lightweight and dynamic convolution methods are proposed for neural machine translation \cite{wu2018pay}. They reduce the parameters by sharing convolution weights between dimensions.
Dynamic convolution is a variant of lightweight convolution where the convolution weight is dynamically predicted through an additional linear layer, which takes a current input feature only. 
Compared to self-attention, they only look at a limited context but performance improvement and faster training compared to Transformer are reported in machine translation.

This paper proposes to apply lightweight and dynamic convolution architectures to E2E ASR and confirms better performance than RNN and performance competitive to state-of-the-art Transformer.
The contributions of this paper are as follows:
\begin{itemize}
\setlength{\itemsep}{-0.1mm}
\setlength{\parskip}{-0.1mm}
 \item Applied lightweight and dynamic convolution methods to E2E ASR for the first time.
 \item Joint training with CTC and lightweight or dynamic convolution.
 \item Additional lightweight or dynamic convolution along the frequency axis.
 \item Combination of different layer types.
 For example, self-attention for the encoder and lightweight convolution for the decoder.
 \item Implement those methods using the open source ASR toolkit ESPnet\cite{Watanabe2018Espnet} to make it publicly available\footnote{https://github.com/espnet/espnet/pull/1599}.
\end{itemize}

There are several prior studies of E2E ASR with convolutional networks. 
In \cite{Zhang2017}, a convolutional layer is applied to the attention-based encoder-decoder model \cite{LAS2016} but still recurrent layers remain.
E2E ASR with only regular convolution is proposed in \cite{zhang2016towards,Neil2018,Li2019}. 
In \cite{Hannun2019}, they use a modified convolution on the encoder side but the decoder is an RNN.
To the best of our knowledge, our work is the first to apply lightweight and dynamic convolution to ASR tasks.

\section{CONVENTIONAL AND PROPOSED METHOD}
\label{sec:method}
This section describes attention-based E2E models by starting from the general formulation. 
Let $X=\{\mathbf{x}_t \in \mathbb{R} ^{d^\text{feat}} | t=1,\cdots,T^\text{feat}\}$ be an acoustic feature sequence and $C=\{c_l \in \mathcal{V} | l=1,\cdots,L \}$ be a token sequence where $d^\text{feat}$ is the dimension of the input feature and $T^\text{feat}$ is its length. $\mathcal{V}$ is a set of distinct tokens and $L$ is the length of a token sequence. Training of the E2E model aims to maximize the following posterior probability:
\begin{equation}
 \label{eq:general_e2e}
 p^\text{e2e}(C|X) = \Pi _{l=1} ^L p(c_l | c_{1:l-1},X).
\end{equation} 
The difference between various E2E models is in how to define the probability distribution in \refeq{general_e2e}.
From the next subsection, the conventional and proposed E2E models are introduced.

\subsection{Attention-based model with RNN}
In the attention-based model with RNN, $p^\text{e2e}(C|X)$ is defined as follows \cite{Shinji2017hybrid,AttentionNIPS2015,LAS2016}.
First, an $\text{Encoder}(\cdot)$ RNN takes an input feature sequence $X$ and outputs the following $d^\text{e}$-dimensional encoded feature sequence $B$:
\begin{equation}
 B = \text{Encoder}(X).
\end{equation}
Here, $B=\{\mathbf{b}_t \in \mathbb{R}^{d^\text{e}} | t=1,\cdots,T^\text{enc}\}$, and $T^\text{enc} \leq T^\text{feat}$ denotes a subsampled input length.
Then,  attention weight $\alpha _{lt}$ and context vector $\mathbf{r}_l$ are calculated as:
\begin{eqnarray}
 \alpha _{lt} = 
    \text{Attention}(\mathbf{q}_{l-1},\mathbf{b}_t),\\
    \mathbf{r}_l = \sum\nolimits _{t=1} ^{T^\text{enc}} \alpha_{lt} \mathbf{b}_t,
\end{eqnarray}
where $\mathbf{q}_l$ is a hidden vector output from $\text{Decoder}(\cdot)$ RNN.
Finally, the probability distribution over output tokens is calculated from $\text{Decoder}(\cdot)$ RNN as:
\begin{equation}
 \label{eq:pdf_of_attention}
 p(c_l | c_{1:l-1},X) = \text{Decoder}(\mathbf{r}_l, \mathbf{q}_{l-1}, c_{l-1}).
\end{equation}
By substituting \refeq{pdf_of_attention} into \refeq{general_e2e}, $p^\text{e2e}(C|X)$ of the attention based model with RNN is defined.
In most cases, long short-term memory (LSTM) RNN is used as $\text{Encoder}(\cdot)$ and $\text{Decoder}(\cdot)$.

\subsection{Transformer}
In contrast to the attention based model with RNN, Transformer does not use any recurrent connection. Instead, self-attention based on multi-head scaled dot product is used as a key component \cite{TransformerNIPS2017}. 
\subsubsection{Multi-head scaled dot product attention}
Scaled dot product attention is defined as:
\begin{equation}
 \text{Attention}(\mathbf{Q}, \mathbf{K}, \mathbf{V}) = \text{Softmax}(\frac{\mathbf{QK}^\text{T}}{\sqrt{d^\text{k}}})\mathbf{V},
\end{equation}
where $\mathbf{Q} \in \mathbb{R}^{T^\text{q} \times d^\text{q}}$, $\mathbf{K} \in \mathbb{R}^{T^\text{k} \times d^\text{k}}$, and $\mathbf{V} \in \mathbb{R}^{T^\text{v} \times d^\text{v}}$ denote query, key, and value matrices, respectively. 
$T^\text{q},T^\text{k}$, and $T^\text{v},$ are the length of each elements and $d^\text{q},d^\text{k}$, and $d^\text{v}$ are the dimensions of each elements.
In Transformer, this attention mechanism is extended to a multi-head one, i.e., 
\begin{align}
 \mathbf{U}_i = \text{Attention}(\mathbf{QW}_i^\text{Q}, \mathbf{KW}_i^\text{K}, \mathbf{VW}_i^\text{V}), \\
 \text{MultiHead}(\mathbf{Q},\mathbf{K},\mathbf{V}) = \text{Concat}(\mathbf{U}_1, \cdots, \mathbf{U}_{H^\text{h}})\mathbf{W}^\text{O}, 
\end{align}
where $H^\text{h}$ is the number of heads. 
The dimension of $\mathbf{Q,K,V}$ are set as $d^\text{q} = d^\text{k} = d^\text{v} = d^{\text{att}}$, $\mathbf{W}_i^\text{Q}, \mathbf{W}_i^\text{K}, \mathbf{W}_i^\text{V} \in \mathbb{R}^{ d^{\text{att}} \times d^{\text{att}}/H^\text{h}}$, and $\mathbf{W}^\text{O} \in \mathbb{R}^{d^{\text{att}} \times d^{\text{att}}}$. 
$d^\text{att}$ is the dimension of $\text{MultiHead}(\cdot)$ input.
Self-attention is the above multi-head attention layer with the scaled dot product whose $\mathbf{Q,K,V}$ are the same:
\begin{equation}
 \text{SelfAttention}(\mathbf{V}) = \text{MultiHead}(\mathbf{V,V,V}).
\end{equation}

\subsubsection{Encoder and decoder of Transformer}
This subsection introduces $p(c_l | c_{1:l-1},X)$ for Transformer.
First, input sequence $X$ is transformed into matrix $\mathbf{X}^\text{E} \in \mathbb{R}^{T^\text{ss} \times d^\text{att}}$ by an input embedding and subsampling operation where $T^\text{SS}$ is the sequence length after subsampling. 
Then, encoded feature $\mathbf{E}$ is calculated as follows:

\begin{equation}
\left\{
\begin{alignedat}{5}
 \mathbf{Z}_{(0)}& =& & \mathbf{X}^\text{E} & & + \mathbf{P}, \\
 \label{eq:enc_sa}
 \mathbf{Z}^\text{A}_{(n)} & =& & \mathbf{Z}_{(n)} & & + \text{SelfAttention}(\mathbf{Z}_{(n)}),\\
 \mathbf{Z}_{(n+1)} & =& & \mathbf{Z}^\text{A}_{(n)} & &+ \text{FF}_n(\mathbf{Z}^\text{A}_{(n)}),\\
 \mathbf{E} & =& & \mathbf{Z}_{(N)} & &,
\end{alignedat}
\right.
\end{equation}
where $n=1,\cdots,N-1$ is the index of encoder layer and $N$ is the layer number.
$\mathbf{P} \in \mathbb{R}^{T^\text{ss} \times {d^\text{att}}} $ is the positional encoding, defined as:
\begin{align}
\left\{
\begin{alignedat}{2}
  & p_{i,2j} & = 
    \sin (i/10000^{2j/d^\text{att}}) ,  \\
  & p_{i,2j+1} & =   
    \cos (i/10000^{2j/d^\text{att}})  .
\end{alignedat}
\right.
\end{align}
$\text{FF}_n(\cdot)$ is defined as:
\begin{equation}
 \text{FF}_n(\mathbf{Z}^\text{A}_{(n)}) = \text{ReLU}(\mathbf{Z}^\text{A}_{(n)} \mathbf{W}_{n1} + \mathbf{b}_{n1})\mathbf{W}_{n2} + \mathbf{b}_{n2},
\end{equation}
where 
$\mathbf{W}_{n1} \in \mathbb{R}^{d^\text{att} \times d^\text{ff}}$, 
$\mathbf{W}_{n2} \in \mathbb{R}^{d^\text{ff} \times d^\text{att}}$, 
$\mathbf{b}_{n1} \in \mathbb{R}^{d^\text{ff}}$,
$\mathbf{b}_{n2} \in \mathbb{R}^{d^\text{att}}$
are parameters to be learned.

Next, $(c_1,\cdots,c_{l-1})$ is transformed to a real-valued matrix $\mathbf{C}^\text{E} \in \mathbb{R}^{L \times d^\text{att}}$ by another input embedding.
Decoder takes $\mathbf{E}$ and $\mathbf{C}^\text{E}$ and calculates 
$[p(c_2 | c_1,X), \cdots, p(c_l | c_{1:c_{l-1}},X)]$ as follows.
\begin{equation}
\left\{
\begin{alignedat}{5}
 \mathbf{Y}_{(0)} & = & & \mathbf{C}^\text{E} & & + \mathbf{P}, \\
 \label{eq:dec_sa}
 \mathbf{Y}^\text{A}_{(m)} & = & & \mathbf{Y}_{(m)} & & + \text{SelfAttention}(\mathbf{Y}_{(m)}),\\
 \mathbf{Y}^\text{S}_{(m)} & = & & \mathbf{Y}^\text{A}_{(m)} & & + \text{MultiHead}(\mathbf{Y}^\text{A}_{(m)},\mathbf{E},\mathbf{E}), \\
 \mathbf{Y}_{(m+1)} & = & & \mathbf{Y}^\text{S}_{(m)} & & + \text{FF}_m(\mathbf{Y}^\text{S}_{(m)}), 
\end{alignedat}
\right.
\end{equation}
\begin{equation}
[p(c_2 | c_1,X), \cdots, p(c_l | c_{1:c_{l-1}},X)] = 
 \text{Softmax}(\mathbf{Y}_{(M)}\mathbf{W}^\text{fin} + \mathbf{b}^\text{fin}).
\end{equation}
$m=1,\cdots,M-1$ is the layer index of the decoder and $M$ is the layer number.
$\mathbf{W}^\text{fin} \in \mathbb{R}^{d^\text{att} \times d^\text{char}}$ and $\mathbf{b}^\text{fin} \in \mathbb{R}^{d^\text{char}}$ are parameters to be learned and $d^\text{char}$ is the number of distinct tokens.

\subsection{Lightweight and dynamic convolution}
Lightweight and dynamic convolution proposed in \cite{wu2018pay} just replaces the self-attention layer of Transformer.
That is, replacing $\text{SelfAttention}(\cdot)$
which takes only value $\mathbf{V}$ as argument  
in \refeq{enc_sa} and \refeq{dec_sa} with 
$\text{LConvLayer}(\cdot)$ or $\text{DConvLayer}(\cdot)$ introduced in this subsection.

As explained in Sec.\ref{sec:intro}, lightweight convolution $(\text{LConv}(\cdot))$ reduces the number of parameters by sharing weights:
\begin{equation}
    \label{eq:lconv}
  \left [\text{LConv}(\mathbf{V},\mathbf{W}^\text{L})\right]_{i,j}
  = \sum _{k=1} ^K w^\text{L}_{\lceil jH^\text{S}/d^\text{v} \rceil,k} \times
  v_{\left( i+k - \lceil \frac{K+1}{2} \rceil \right), j},
\end{equation}
where $\mathbf{W}^\text{L} \in \mathbb{R}^{H^\text{S} \times K}$ is a convolution kernel to be learned and $H^\text{S}$ is a weight sharing parameter. Compared to depthwise convolution, the number of parameters is reduced from $d^\text{v} K$ to $H^\text{S}K$ by setting $H^\text{S}<d^\text{v}$.

Dynamic convolution $(\text{DConv}(\cdot))$ uses a kernel estimated from the current input:
\begin{equation}
\label{eq:dconv}
\text{DConv}(\mathbf{V}) = \text{LConv}(\mathbf{V}, \mathbf{W}^\text{D} \mathbf{v}_t),
\end{equation}
where $\mathbf{W}^\text{D} \in \mathbb{R}^{ H^\text{S} \times K \times d^\text{v} }$ is a parameter to be learned.

Then, $\text{LConvLayer}(\cdot)$ and $\text{DConvLayer}(\cdot)$ are defined as follows:
\begin{align}
 \text{LConvLayer}(\mathbf{V}) & = 
 \text{LConv}(\text{GLU}(\mathbf{VW}^\text{I}),\mathbf{W}^\text{L})\mathbf{W}^\text{P}, \\
 \text{DConvLayer}(\mathbf{V}) & = 
 \text{DConv}(\text{GLU}(\mathbf{VW}^\text{I}))\mathbf{W}^\text{P},
\end{align}
where $\mathbf{W}^\text{I} \in \mathbb{R}^{d^\text{v} \times 2d^\text{v}}$ and $\mathbf{W}^\text{P} \in \mathbb{R}^{d^\text{v} \times d^\text{v}}$ are parameters to be learned. $\text{GLU}(\cdot)$ is a gated linear unit \cite{dauphin2017language}.

\begin{table}[tp]
 \begin{center}
  \small
 \begin{tabular}{|c|c|c|}
  Model ID & Encoder & Decoder \\ \hline
  SA & \text{SelfAttention} & \text{SelfAttention} \\
  LC & \text{LConvLayer} & \text{LConvLayer} \\
  DC & \text{DConvLayer} & \text{DConvLayer} \\
  LC2D & \text{LConv2DLayer} & \text{LConv2DLayer} \\
  DC2D & \text{DConv2DLayer} & \text{DConv2DLayer} \\
  SA-LC & \text{SelfAttention} & \text{LConvLayer} \\
  SA-DC & \text{SelfAttention} & \text{DConvLayer} \\
  SA-LC2D & \text{SelfAttention} & \text{LConv2DLayer} \\
  SA-DC2D & \text{SelfAttention} & \text{DConv2DLayer} \\
 \end{tabular}
  \caption{
    \label{tb:layer_comb}
    List of all combinations of layers for encoder and decoder used in experiments.
    }
 \end{center}
\end{table}

\subsubsection{Adding convolution along the frequency axis}
We propose to add convolution along the frequency axis to lightweight and dynamic convolution.
This is inspired by a variant of LSTMs that perform recurrence along the additional frequency axis \cite{Sainath+2016,li2015lstm}.

In case of lightweight convolution the frequency axis convolution is performed in contrast to \refeq{lconv} as:
\begin{align}
 \left [ \text{LConvF}(\mathbf{V},\mathbf{w}^\text{F})\right]_{i,j} = \sum_{k=1}^K w_k^\text{F} \times v_{i, \left({j+k - \lceil \frac{K+1}{2} \rceil}\right)},
\end{align}
where 
$\mathbf{w}^\text{F} \in \mathbb{R}^{K}$ 
are kernel weights to be estimated during training.

Similarly to \refeq{dconv}, dynamic convolution along the frequency axis is performed as:
\begin{equation}
\text{DConvF}(\mathbf{V}) = 
 \text{LConvF}(\mathbf{V},\mathbf{W}^\text{U} \mathbf{v}_t),
\end{equation}
where
$\mathbf{W}^\text{U} \in \mathbb{R}^{K \times d^\text{v}}$ is a learnable parameter.

Then, these outputs are concatenated with the output of lightweight or dynamic convolution. Finally, $\text{LConv2DLayer}(\cdot)$ and \\ $\text{DConv2DLayer}(\cdot)$ are defined as:
\begin{align}
 & \text{LConv2DLayer}(\mathbf{V}) \nonumber \\
 & \quad = 
 \text{Concat}
  \left(
    \text{LConv}(\mathbf{G},\mathbf{W}^\text{L}),  
    \text{LConvF}(\mathbf{G},\mathbf{w}^\text{F}) 
  \right)\mathbf{W}^\text{R}, \\
 & \text{DConv2DLayer}(\mathbf{V}) \nonumber \\
 & \quad = \text{Concat}
  \left(
    \text{DConv}(\mathbf{G}), 
    \text{DConvF}(\mathbf{G}) 
  \right)\mathbf{W}^\text{R},
\end{align}
where  $\mathbf{W}^\text{R} \in \mathbb{R}^{2d^\text{v} \times d^\text{v}}$ is a parameter to be learned and $\mathbf{G} = \text{GLU}(\mathbf{VW}^\text{I})$.


\begin{table*}[tb]
 \begin{center}
  \small
 \begin{tabular}{cccc|rrr|rrr|rrrr}
\hline \hline 
  \multicolumn{10}{c}{}   &  \multicolumn{4}{|c}{\textit{\textbf{Librispeech}}} \\
  \multicolumn{4}{c}{} & \multicolumn{3}{|c|}{\textit{\textbf{CSJ 271h}}} & \multicolumn{3}{c}{\textit{\textbf{CSJ 581h}}} & \multicolumn{2}{|c}{\textit{\textbf{dev}}} & \multicolumn{2}{c}{\textit{\textbf{test}}} \\
  Model ID & $H^\text{S}$ & $K^\text{e}$ & $K^\text{d}$  & \textit{\textbf{eval1}} & \textit{\textbf{eval2}} & \textit{\textbf{eval3}} & \textit{\textbf{eval1}} & \textit{\textbf{eval2}} & \textit{\textbf{eval3}} & \textit{\textbf{clean}} & \textit{\textbf{other}} & \textit{\textbf{clean}} & \textit{\textbf{other}}\\ \hline \hline
  RNN & N.A. & N.A & N.A & 8.5 & 6.3 & 15.9 & 6.5 & 4.8 & 5.1 & 4.0 & 12.0 & 4.1 & 12.8 \\
  SA & N.A. & N.A & N.A & 7.1 & 5.0 & 12.6 & {\bf 5.5} & 3.9 & 4.3 & 3.7 &  {\bf 9.6} & 3.9 &  9.8 \\ \hline
  LC & 4 & 101 & 71 & 7.6 & 5.3 & 13.0 & 5.9 & 4.2 & 4.6 & 3.5 & 10.2 & 3.7 & 10.7\\
  DC & 4 & 101 & 71 & 7.9 & 5.5 & 13.5 & 6.2 & 4.2 & 4.5 & 3.5 & 10.5 & {\bf 3.6} & 10.8 \\
  LC2D & 16 & 101 & 71 & 7.6 & 5.4 & 12.8 & 5.8 & 4.1 & 4.4 & {\bf 3.4} & 10.3 & 3.7 & 10.6 \\
  DC2D & 2 & 31 & 11 & 8.2 & 5.8 & 13.2 & 6.5 & 4.5 & 4.7 & 3.6 & 11.5 & 3.8 & 11.6 \\ \hline
  SA-LC & 8 & N.A. & 31 & {\bf 7.0} & {\bf 4.9} & 12.6 & 5.6 & 4.1 & 4.3 & 3.8 & {\bf 9.6} & 4.2 & 9.8 \\ 
  SA-DC & 8 & N.A. & 31 & 7.1 & 5.0 & {\bf 12.3} & 5.6 & {\bf 3.8} & {\bf 4.1} & 4.2 & 9.9 & 4.6 & 10.2 \\
  SA-LC2D & 4 & N.A. & 11 & 7.1 & {\bf 4.9} & 12.5 & {\bf 5.5} & 4.1 & 4.2 & 3.9 & {\bf 9.6} & 4.3 & 9.7 \\ 
  SA-DC2D & 4 & N.A. & 11 & {\bf 7.0} & {\bf 4.9} & 12.6 & 5.6 & 4.0 & {\bf 4.1} & 3.5 & {\bf 9.6} & 3.9 & {\bf 9.6} \\ \hline \hline 
 \end{tabular}
  \caption{
    \label{tb:results}
    Character error rate (CER) of CSJ and word error rate (WER) of Librispeech.
    Using self-attention for encoder and convolutional layer for decoder (SA-\{LC,DC,LC2D,DC2D\} in \reftb{layer_comb}) yield better performance than RNN and performance competitive to state-of-the-art Transformer. (SA in \reftb{layer_comb}).
  }
 \end{center}
\end{table*}

\begin{table*}[tb]
 \begin{center}
  \small
 \begin{tabular}{cccc|rrrrrr|rr|rrrr}
    \hline \hline
  \multicolumn{4}{c}{}   & \multicolumn{6}{|c|}{\textit{\textbf{REVERB Simulated}}} & \multicolumn{2}{c|}{\textit{\textbf{REVERB Real}}}  &  \multicolumn{4}{c}{\textit{\textbf{CHiME4}}}\\
   \multicolumn{4}{c}{}   & \multicolumn{2}{|c}{\textit{\textbf{Room 1}}} & \multicolumn{2}{c}{\textit{\textbf{Room 2}}} & \multicolumn{2}{c|}{\textit{\textbf{Room 3}}} & \multicolumn{2}{c|}{\textit{\textbf{Room 1}}} & \multicolumn{2}{c}{\textit{\textbf{dev}}} & \multicolumn{2}{c}{\textit{\textbf{test}}}\\
  Model ID & $H^\text{S}$ & $K^\text{e}$ & $K^\text{d}$  & \textit{\textbf{Near}} & \textit{\textbf{Far}} & \textit{\textbf{Near}} & \textit{\textbf{Far}} & \textit{\textbf{Near}} & \textit{\textbf{Far}} & \textit{\textbf{Near}} & \textit{\textbf{Far}} & \textit{\textbf{simu}} & \textit{\textbf{real}} & \textit{\textbf{simu}} & \textit{\textbf{real}} \\ \hline \hline
  RNN & N.A. & N.A & N.A & 5.7 & 5.7 & 5.7 & 5.9 & 5.9 & 6.3 & 18.4 & 19.2 & 10.4 & 9.9 & 20.4 & 19.2 \\ 
  SA & N.A. & N.A & N.A & 5.6 & 6.1 & 6.2 & 6.4 & 5.9 & 6.1 & 10.9 & 13.7 & 10.1 & 8.9 & 16.9 & 15.8 \\ \hline
  LC & 4 & 101 & 71 & 6.3 & 6.5 & 6.7 & 6.5 & 6.5 & 7.0 & 12.7 & 14.0& 10.5 & 9.1 & 18.4 & 16.5\\
  DC & 4 & 101 & 71 & 6.5 & 6.7 & 6.8 & 7.0 & 7.1 & 6.8 & 16.2 & 17.0& 11.0 & 8.7 & 19.1 & 17.4 \\
  LC2D & 16 & 101 & 71 & 5.1 & 5.9 & 5.8 & 5.8 & 5.6 & 6.1 & 11.0 & 13.1 & 10.1 & {\bf 8.2} & 17.5 & 16.0 \\
  DC2D & 2 & 31 & 11 & 6.0 & 6.1 & 6.2 & 6.2 & 6.1 & 6.9 & 16.0 & 16.9 &11.6& 10.1 &21.6& 19.4 \\ \hline
  SA-LC & 8 & N.A. & 31 & 4.7 & 4.9 & {\bf 4.6} & 4.9 & 4.8 & 5.4 & 8.5 & 11.8& 9.6 & 8.6 & 16.3 & 16.1\\ 
  SA-DC & 8 & N.A. & 31 & 4.5 & 4.9 & 4.9 & 5.1 & 5.3 & 5.3 & 9.3 & 12.4 & {\bf 9.3} & {\bf 8.2} & 16.7 & 16.0 \\
  SA-LC2D & 4 & N.A. & 11 &  {\bf 4.4} & 4.7 & 4.7 & {\bf 4.8} & {\bf 4.7} & {\bf 5.2} & 8.0 & {\bf 12.0}& 9.5 & 8.3 & 16.7 & 15.4 \\ 
  SA-DC2D & 4 & N.A. & 11 & 4.5 & {\bf 4.5} & 4.9 & 5.0 & 5.1 & 5.3 & {\bf 7.5} & 12.1 & 9.6 & 8.3 & {\bf 16.2} & {\bf 15.3}\\ \hline \hline 
 \end{tabular}
  \caption{
    \label{tb:results2}
    Word error rate (WER) of REVERB and CHiME4. 
    Using self-attention for encoder and convolutional layer for decoder (SA-\{LC,DC,LC2D,DC2D\} in \reftb{layer_comb}) yield better performance than RNN and performance competitive to state-of-the-art Transformer.
  }
 \end{center}
\end{table*}

\subsection{Combination of different layers}
We also propose using a different layer to $\text{SelfAttention}(\cdot)$ between the encoder and decoder of Transformer.
All combinations that we tried are summarized in \reftb{layer_comb}. Using $\text{SelfAttention}(\cdot)$ as decoder and convolutional layer as encoder do not work well according to preliminary experiments so has been omitted.

\subsection{Joint training and decoding with CTC}
Same as in \cite{Shinji2017hybrid,Karita2019}, joint training and decoding with CTC objective is used. 
Linear transformation and Softmax are applied to the encoded feature $\mathbf{E}$ then fed to CTC loss. The CTC loss is interpolated with $p^\text{e2e}$ and used as the objective function. 
CTC helps the attention-based decoder to learn monotonic alignment which results in faster convergence and performance improvement.

\section{EXPERIMENTS}
\label{sec:exp}
\subsection{Setup}
In order to evaluate our proposed method, we used four corpora, CSJ \cite{CSJ2000}, Librispeech \cite{Libri2015}, REVERVB \cite{Kinoshita2016} and CHiME4 \cite{chime4}. 
As a baseline, the CTC/Attention hybrid method proposed in \cite{Shinji2017hybrid} was used. The default recipe of ESPnet toolkit \cite{Watanabe2018Espnet} was used. We call this baseline RNN.

Another baseline was Transformer proposed in \cite{Karita2019}. The recipe published by the author of ESPnet was used ($N=12, M=6, d^\text{att}=256, H^\text{h}=4$) but gradient accumulation was increased to 8 because it resulted in better performance. This corresponds to SA in \reftb{layer_comb}.  

For REVERB, single channel training with all the 8 channel simulation data was used. 
During evaluation the 8 channel data are processed by frontend of weighted prediction error (WPE) \cite{wpe} followed by delay-and-sum beamforming (BeamformIt) \cite{anguera2007acoustic}. 
For CHiME4, again single channel training with all the 6 channel data was used. 
For evaluation, the 5 channel data was used (second channel was excluded) with BeamformIt as frontend.

For lightweight and dynamic convolutions and our proposed architecture, basically the same recipe for Transformer mentioned above was used. The differences were batch size and gradient accumulation used for DC. Batch size was set to 48 and gradient accumulation was 11 batches in order to stabilize training. Also, DropConnect \cite{wu2018pay} with probability 0.1 was used. The recipes we used will be made publicly available in the ESPnet repository.

\subsection{Results}
We first looked for the best hyper parameter such as kernel length of encoder $K^\text{e}$ and decoder $K^\text{d}$ and weight sharing $H^\text{S}$ with small amount of training data. Note that $K^\text{e}$, $K^\text{d}$ and $H^\text{S}$ are the same for all layers.
Character error rate (CER) of CSJ trained with only Academic lecture data (271 hours) is shown in \reftb{results}. We picked the best hyperparameter for each model while keeping the total number of parameters almost the same. 
Because of space limitation not all experiments are shown but larger kernel length led to better CER while smaller $H^\text{S}$ had limited impact.
By combining self-attention and convolutional layer, CER is better than RNN and is competitive to state-of-the-art Transformer.
Adding convolution along the frequency axis has little effect on CER.

With the best hyperparameter obtained above, we evaluated our methods with a large amount of training data and noisy/reverberant data.
CER of CSJ (581 hours) and word error rate (WER) of Librispeech (960 hours) are shown in \reftb{results}. 
WER of REVERB and CHiME4 are shown in \reftb{results2}. 
For all the test sets, the combination of self-attention and convolutional layer achieved better performance than RNN and performance competitive to state-of-the-art Transformer. 
For clean sets of Librispeech, models with only a convolutional layer yield the best performance.


\subsection{Discussion}
  By using lightweight or dynamic convolution for both encoder and decoder, performance is comparable to, or better than, RNN except for some sets of REVERB and CHiME4.
  For clean sets of Librispeech, performance is even better than Transformer.
  This means the convolutional layer, which looks only at a limited context, is more suitable for E2E-ASR of clean speech than RNN which encodes the whole utterance.

  By using self-attention to encoder and convolutional layer to decoder, performance for all test sets is reaching to, or better than, state-of-the-art Transformer. Especially for REVERB and CHiME4, performance gains are larger than on other corpora. 
  This means, at least for the decoder side, a convolutional layer is 
  better probably because the limited context could eliminate the effect of the wrong recognition history caused by noisy speech to be maintained for the entire utterance during beam search.
  
  The effectiveness of adding convolution along the frequency axis is dependent on the test set.
  For example, on all sets of Librispeech, test sets of CHiME4 and real sets of REVERB, SA-DC2D is better than SA-DC. However, on another sets it is not always the case.
  Further analysis is necessary to conclude the effectiveness of adding convolution along the frequency axis.


\section{CONCLUSION}
\label{sec:conc}
In this paper lightweight and dynamic convolution originally proposed for machine translation is applied to ASR. In theory, training is faster because computational complexity is linear in input length while Transformer is quadratic. We also propose to use joint training and decoding with the CTC objective, convolution along the frequency axis and combination of self-attention and convolutional layer to achieve better performance. 
ASR experiments on various corpora showed our proposed method yields better performance than RNN and performance competitive to state-of-the-art Transformer. 
Detailed analysis of the results and computational efficiency are left as future work.

\vfill\pagebreak

\bibliographystyle{IEEEbib}
\bibliography{refs}

\end{document}